\documentclass[a4paper,showpacs,prb,twocolumn]{revtex4}
\usepackage{amsfonts}
\usepackage{graphicx}
\usepackage{color}
\usepackage{amsmath}
\usepackage{amssymb}
\usepackage{latexsym}

\begin{document}

\title{Many-body effects in transport through open systems: pinning of
resonant levels}
\author{S. Ihnatsenka}
\affiliation{Solid State Electronics, Department of Science and Technology (ITN), Link%
\"{o}ping University, 60174 Norrk\"{o}ping, Sweden}
\author{I. V. Zozoulenko}
\affiliation{Solid State Electronics, Department of Science and Technology (ITN), Link%
\"{o}ping University, 60174 Norrk\"{o}ping, Sweden}
\author{M. Willander}
\affiliation{Solid State Electronics, Department of Science and Technology (ITN), Link%
\"{o}ping University, 60174 Norrk\"{o}ping, Sweden}
\affiliation{Department of Physics, G\"{o}teborg University, 412 96 G\"{o}teborg, Sweden}
\date{\today }

\begin{abstract}
The role of electron-electron interaction in transport properties of open
quantum dots is studied. The self-consistent full quantum mechanical
magnetotransport calculations within the Hartree, Density Functional Theory
and Thomas-Fermi approximations were performed where a whole device,
including the semi-infinitive leads, is treated on the same footing (i.e.
the electron-electron interaction is accounted for both in the leads as well
as in the dot region). The main finding of the present paper is the effect
of pinning of the resonant levels to the Fermi energy due to the enhanced
screening. Our results represent a significant departure from a conventional
picture where a variation of external parameters (such as a gate voltage,
magnetic field, etc.) causes the successive dot states to sweep past the
Fermi level in a linear fashion. We instead demonstrate highly nonlinear
behavior of the resonant levels in the vicinity of the Fermi energy. The
pinning of the resonant levels in open quantum dots leads to the broadening
of the conduction oscillations in comparison to the one electron picture.
The effect of pinning becomes much more pronounced in the presence of the
perpendicular magnetic field. This can be attributed to the enhanced
screening efficiency because of the increased localization of the wave
function. The strong pinning of the resonant energy levels in the presence
of magnetic field can have a profound effect on transport properties of
various devices operating in the edge state transport regime. We also
critically examine an approximation often used in transport calculations
where an inherently open system is replaced by a corresponding closed one.
\end{abstract}

 \maketitle

\section{Introduction}

A transport regime where a sub-micron lateral structure is strongly coupled
to electron reservoirs (leads) is usually referred to as an open one\cite%
{Alhassid,Datta}. This transport regime can be realized in quantum wires,
dot and antidot structures typically fabricated using a split-gate or
related techniques. Such the techniques allow one to obtain devices with
desired and variable geometry and parameters such as an electron density and
lead openings. The quantum dot operates in an open regime when the gate
voltage sets up two quantum point contacts (QPCs) at the entrance and exit
of the dot such that they transmit one or more channels (i.e. the
conductance of an individual QPC, $G_{QPC}\gtrsim \frac{2e^{2}}{h}$). In
this regime electrons can freely enter and exit the dot, such that the
electron number inside the dot is not integer and the chemical potential
throughout the whole device in the linear responce regime is constant.
[Opposite regime emerges when the point contacts are nearly pinched off and
is referred to as a Coulomb blockade\cite{Datta,Kastner}. In this case the
electron number in the dot is quantized and the chemical potential inside
the dot is different from that one in the leads]. During the last decade the
open quantum dots have received a significant attention providing many
important insights into areas such as quantum interference, chaos,
decoherence, localization and many others\cite{Alhassid}. Earlier transport
experiments have been mainly devoted to the dots with hundreds or even
thousands electrons. Only recently it has become possible to reduce
occupancy down to only a few or even one electron\cite{Ciorga,Zozoulenko_PRL}%
.

Electron-electron interaction is known to have great impact on transport in
quantum dots with such pronounced examples as Coulomb blockade\cite{Kastner}
or Kondo effect\cite{Kondo}. 
A description of the quantum transport in quantum dots is often based on
model Hamiltonians containing phenomenological parameters such as coupling
strengths or charging constants\cite{Henrickson,Indlekofer,EurophLett}. In
many cases it is not always straightforward to relate quantitatively the
above parameters to the physical processes they represent in the real system
and sometimes it is not even obvious whether a model description is
sufficient to capture the essential physics. At the same time, it is now
becomes well recognized that a detailed understanding and interpretation of
the experiment might require a quantitative microscopical modelling of the
system at hand free from phenomenological parameters and not relying on
model Hamiltonians which validity is poorly controlled. The importance of
such the modelling can be illustrated by examples including the quantitative
description of the compressible/incompressible strips in magnetic field at
the edges of the two-dimensional electron gas\cite{Chklovskii} or
explanation of the Hund rule observed in few-electron quantum dots\cite%
{Stephanie}, just to name a few.

The purpose of the present paper is twofold. First, we develop an approach
aimed on full quantum mechanical many-body transport calculations in open
systems that starts from the lithographical layout of the device and does
not include phenomenological parameters such as coupling strengths, charging
constants etc. The whole device, including semi-infinitive leads, is treated
on the same footing (i.e. the electron-electron interaction is accounted for
both in the leads as well as in the dot region). Using the recursive Green's
function technique we self-consistently compute the scattering solutions of
the two-dimensional Schr\"{o}dinger equation in magnetic field\cite%
{Zozoulenko_1996}. Following the parametrization for the exchange and
correlation energy functionals of Tanatar and Ceperley\cite{TC} the
electron-electron interaction is incorporated within the density functional
theory (DFT) in the local density approximation\cite{ParrYang,Stephanie}.
The validity of the DFT approximation is supported by the excellent
agreement with the exact diagonalization and variational Monte Carlo
calculations performed for few-electron systems\cite{validity}, as well as
by the good quantitative correspondence between the experiment and the DFT
calculations for the magnetoconductance of quantum wires\cite{Radu}. Our
approach thus accounts for the quantum mechanics nature of the scattering
states and the resonant levels as well as for the exchange and correlation
beyond the Hartree approximation. At the same time, it accurately describes
the global electrostatic and the screening in the dots as well as in the
leads.

The second aim of the present paper is to revise the role of the
electron-electron interaction in transport properties of open quantum dots.
It is widely believed that in open transport regime (as opposed to the
Coulomb blockade or Kondo regime), the electron-electron interaction plays
only a minor role. The main finding of the present paper is the effect of
pinning of the resonant levels to the Fermi energy due to the enhanced
screening. Our results represent a significant departure from a conventional
picture adopted in most model Hamiltonians as well as in more sophisticated
numerical calculations where a variation of external parameters (such as a
gate voltage, magnetic field, etc.) causes the successive dot states to
sweep past the Fermi level in a linear fashion. We instead demonstrate
highly nonlinear behavior of the resonant levels in the vicinity of the
Fermi energy. One of the observable consequence of this effect is smearing
of the conductance fluctuations. We also show that the resonant level
pinning becomes especially pronounced in magnetic field. Thus, accounting
for this effect might be important for the interpretation of the
magnetotransport experiment in open structures, including e.g. recent
studies of the electronic Mach-Zehnder interferometer\cite{Mach-Zender} and
the Laughlin quasiparticle interferometer\cite{Goldman}, structures designed
to test the realization of the topological quantum computing\cite{QuantComp}%
, antidot structures\cite{adot,Karakurt,APL} and others. It should be also
noted that the quantum dot structures demonstrating Kondo effect fall into
the semi-open transport regime such that accounting for effect of the
nonlinear screening leading to the resonant level pinning might be essential
for the interpretation of the experiments in this regime as well.


The paper is organized as follow. Section II presents the model and the
Hamiltonian of the system at hand. In Sec. III the numerical method for the
self-consistent calculation of the magnetoconductance is described.
Computational results are presented and discussed in Sec. IV, and
conclusions are given in Sec. V.

\section{Model}

\begin{figure}[!tb]
\includegraphics[scale=0.5]{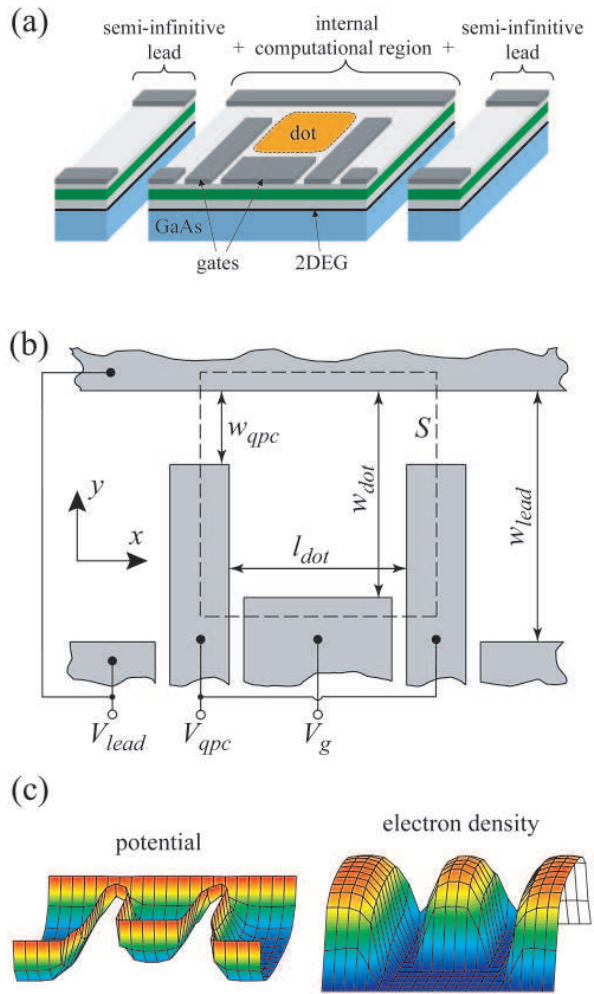}
\caption{(color online) (a) Structure of an open quantum dot. The internal
region is attached to two semi-infinitive quantum wires which serve as the
electron reservoirs. (b) The layout of the gates defining the dot. The
dashed line defines the area $S$ of the quantum dot used to calculate a
number of electrons in it. (c) Representative calculated the self-consistent
potential and the electron density.}
\label{f:structure}
\end{figure}

We consider an open quantum dot attached to semi-infinitive leads (electron
reservoirs) in a perpendicular magnetic field $B$. A schematic layout of the
device is illustrated in Fig. \ref{f:structure}(a). Charge carriers
originating from a fully ionized donor layer form the two-dimensional
electron gas (2DEG) which is buried inside a substrate at the GaAs/Al$_{x}$Ga%
$_{1-x}$As heterointerface situated at the distance $b$ from the surface.
Metallic gates placed on the top of the heterostructure define the dot and
the leads on the depth of the 2DEG (Figs. \ref{f:structure} (a),(b)).

The Hamiltonian of the whole system (the dot + the leads) can be written in
the form
\begin{equation}
H=H_{0}+V(\mathbf{r})  \label{Hamiltonian}
\end{equation}%
where $H_{0}$ is the kinetic energy in the Landau gauge, $\mathbf{A}%
=(-By,0,0)$,
\begin{equation}
H_{0}=-\frac{\hbar ^{2}}{2m^{\ast }}\left\{ \left( \frac{\partial }{\partial
x}-\frac{eiBy}{\hbar }\right) ^{2}+\frac{\partial ^{2}}{\partial y^{2}}%
\right\} ,  \label{2}
\end{equation}%
$\mathbf{r}=(x,y),$ $m^{\ast }=0.067m_{e}$ is the GaAs effective mass. The
total confining potential within the framework of the density functional
theory is the sum of the electrostatic confinement potential, the Hartree
potential, and the exchange-correlation potential
\begin{equation}
V(\mathbf{r})=V_{conf}(\mathbf{r})+V_{H}(\mathbf{r})+V_{xc}(\mathbf{r}).
\label{V_tot}
\end{equation}%
The electrostatic confinement $V_{conf}(\mathbf{r})=V_{gates}(\mathbf{r}%
)+V_{donors}+V_{Schottky}$ includes contributions respectively from the top
gates, the donor layer and the Schottky barrier. The explicit expressions
for the potentials $V_{gates}(\mathbf{r})$ and $V_{donors}$ are given in
respectively Ref. \onlinecite{Davies_gate} and Ref. %
\onlinecite{Martorell_donor}; the Schottky barrier is chosen to be $%
V_{Schottky}=0.8$ eV. The Hartree potential is written in a standard form%
\begin{equation}
V_{H}(\mathbf{r})=\frac{e^{2}}{4\pi \varepsilon _{0}\varepsilon _{r}}\int d%
\mathbf{r}\,^{\prime }n(\mathbf{r}^{\prime })\left( \frac{1}{|\mathbf{r}-%
\mathbf{r}^{\prime }|}-\frac{1}{\sqrt{|\mathbf{r}-\mathbf{r}^{\prime
}|^{2}+4b^{2}}}\right) ,  \label{V_H}
\end{equation}%
where $n(\mathbf{r})$ is the electron density and the second term describes
the mirror charges places at the distance of $b$ from the surface, $n(%
\mathbf{r})$ is the electron density, $\varepsilon _{r}=12.9$ is the
dielectric constant of GaAs, and the integration is performed over the whole
device area including the semi-infinite leads.

The last term in the total confining potential (\ref{V_tot}) is the exchange
and correlation potential $V_{xc}[n(\mathbf{r})]=V_{x}[n(\mathbf{r}%
)]+V_{c}[n(\mathbf{r})]$ which is the functional of the electron density. In
the local density approximation it is given by a functional derivative\cite%
{ParrYang}
\begin{equation}
V_{xc}=\frac{d}{dn}\left\{ n\epsilon _{xc}\left( n\right) \right\} .
\label{LDA}
\end{equation}%
For $\epsilon _{xc}$ we have used the parametrization of Tanatar and
Ceperley \cite{TC} for the case of spin degenerate electrons. In particular,
for the exchange potential this parametrization gives $V_{x}[n(\mathbf{r})]=-%
\frac{e^{2}}{\varepsilon _{0}\varepsilon _{r}\pi ^{3/2}}\sqrt{n(\mathbf{r})/2%
}$. Note that setting $V_{xc}(\mathbf{r})=0$ in Eq. (\ref{V_tot}) we reduce
our approach to the standard Hartree approximation.

To outline the role of quantum mechanical effects in the electron-electron
interaction in open quantum dots we also consider the Thomas-Fermi (TF)
approximation. In this approximation the kinetic energy is related to the
electron density \cite{ParrYang}, $H_{0}=\frac{\pi \hbar ^{2}}{m^{\ast }}n(%
\mathbf{r})$. The self-consistent electron density is thus obtained from the
solution of the equation%
\begin{equation}
\frac{\pi \hbar ^{2}}{m^{\ast }}n(x,y)+V_{conf}(r)+V_{H}(r)=E_{F}.
\label{TF}
\end{equation}%
The electron density and the total confining potential calculated within the
TF approximation do not capture quantum-mechanical quantization of the
electron motion. The utilization of the TF approximation for the modelling
of the magnetotransport in open system is therefore conceptually equivalent
to a one-electron approach. The difference between these approaches is the
shape of the total confining potential: in one-electron transport
simulations one typically starts with a model hard-wall confinement, whereas
the TF approximation gives a rather smooth potential which represents a good
approximation to the actual confinement.

\section{Method}

\begin{figure}[tb]
\includegraphics[scale = 0.6]{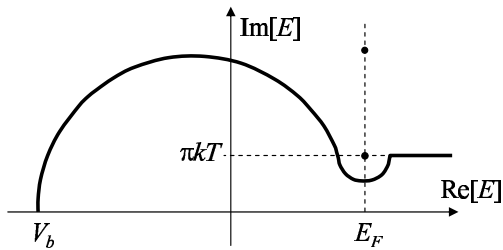}
\caption{A typical integration contour used in the calculation of integral (%
\protect\ref{density}). Dots indicate the poles of the Fermi-Dirac
distribution function in the upper complex plane at $\Re \left[ E\right]
=E_{F}$, $\Im \left[ E\right] =(2m+1)\protect\pi kT,$ $m=0,1,2,\ldots $ . $%
V_{b}$ is the bottom of the conduction band in the leads (the lowest
potential in the system).}
\label{f:contour}
\end{figure}

The magnetoconductance through the quantum dot in the linear response regime
is given by the Landauer formula\cite{Datta_book}
\begin{equation}
G=-\frac{2e^{2}}{h}\int dE\,T(E)\frac{\partial f_{FD}\left( E-E_{F}\right) }{%
\partial E},  \label{conductance}
\end{equation}%
where $T(E)$ is the total transmission coefficient, $f_{FD}(E-E_{F})$ is the
Fermi-Dirac distribution function and $E_{F}$ is the Fermy energy. In order
to calculate $T(E)$ in perpendicular magnetic field we utilize the recursive
Greens function technique in the hybrid energy-space representation\cite%
{Zozoulenko_1996}. We discretize Eq. (\ref{Hamiltonian}) and introduce the
tight-binding Hamiltonian (with lattice constant $a=4$ nm), where the
perpendicular magnetic field is included in a form of the Peierl's
substitution \cite{Datta_book}. The retarded Green's function is introduced
in a standard way\cite{Datta_book},
\begin{equation}
\mathcal{G}=\left( E-H+i\eta \right) ^{-1}.  \label{gfunction}
\end{equation}%
The Green's function in the real space representation, $\mathcal{G}(\mathbf{r%
},\mathbf{r},E)$, provides an information about the electron density at the
site $\mathbf{r},$ \cite{Datta_book}
\begin{equation}
n(\mathbf{r})=-\frac{1}{\pi }\Im \int dE\,\mathcal{G}(\mathbf{r},\mathbf{r}%
,E)\,f(E-E_{F}).  \label{density}
\end{equation}%
Note that $\mathcal{G}(\mathbf{r},\mathbf{r},E)$ is a rapidly varying
function of energy. As a result, a direct integration along the real axis in
Eq. (\ref{density}) is rather ineffective as its numerical accuracy is not
sufficient to achieve convergence of the self-consistent electron density.
Because of this, we transform the integration contour into the complex plane
$\Im \lbrack E]>0,$ where the Green's function is much more smoother. Note
that all poles of the retarded Green's function are in the lower half-plane $%
\Im \lbrack E]<0$. A typical contour used in the integration avoiding poles
of the Fermi-Dirac function is shown in Fig. \ref{f:contour}.

In order to calculate the Green's function of the whole system (dot + leads)
we divide the system into three parts, the internal computational region and
two semi-infinite leads as shown in Fig. \ref{f:structure} (a). Note that
the internal region incorporates not only the dot, but also the straight
segments including a part of the leads. We place the semi-infinite leads
sufficiently far away from the dot where the total self-consistent potential
and the electron density do not change along the leads (i.e. the electron
density and the potential in the leads are not affected by the internal
region, such that the leads can be considered as uniform quantum wires.).
This allows us to calculate the total potential and the electron density in
the lead regions using the technique developed in Ref. %
\onlinecite{Ihnatsenka} for an infinite homogeneous channel in the
perpendicular magnetic field. The Green's function in the internal region is
calculated using the standard recursive Green's function technique. The
total Green's function for the whole system is calculated by linking with
the help of the Dyson equation the surface Greens function for the
semi-infinite leads (with the self-consistent potential calculated using the
technique of Ref. \onlinecite{Ihnatsenka}) and the Green's function of the
internal region.

All the calculations described above are performed self-consistently in an
iterative way until a converged solution for the electron density and
potential (and hence for the total Green's function) is obtained. Having
calculated the total self-consistent Green's function, the scattering
problem is solved where the scattering states in the leads (both propagating
and evanescent) are obtained using the Green's function technique of Ref. %
\onlinecite{Ihnatsenka} (The equation for the calculation of the
transmission and reflection coefficients using the Green's function in the
presence of the magnetic field is derived in Ref. %
\onlinecite{Zozoulenko_1996}).

Having calculated the Green's function of the internal region and the wave
function in the leads we can recover the wave function $\psi (\mathbf{r},E)$
inside the internal region. For visualization of the wave function inside
the dot we include the effect of the finite temperature as follows,
\begin{equation}
\left\vert \Psi (\mathbf{r})\right\vert ^{2}=-\int dE\,\left\vert \psi (%
\mathbf{r},E)\right\vert ^{2}\frac{\partial f_{FD}\left( E-E_{F}\right) }{%
\partial E}.  \label{psi}
\end{equation}%
In order to find the DOS inside the quantum dot\cite{Datta_book} we perform
integration over the dot area $S$ defined in Fig. \ref{f:structure}(b),
\begin{equation}
\text{DOS}(E)=-\frac{1}{\pi }\Im \int_{S}d\mathbf{r\,}\mathcal{G}(\mathbf{r},%
\mathbf{r},E).  \label{dos}
\end{equation}

Note that a many-body approach conceptually similar to ours for calculation
of the quantum transport in an open dot was developed in Ref. %
\onlinecite{Guo} for the case of zero magnetic field. The magnetic field was
included in the dot region in the transport calculations presented in Ref. %
\onlinecite{Evaldsson} where, however, the leads were considered as
non-interacting.

As we mentioned in the previous section, we also employ the TF approximation
for the calculation of the magnetotransport through the quantum dot. This
calculation is done by the same method as described above, with the only
difference that the self-consistent electron densities in the internal
region and in the semi-infinite leads are calculated from the semi-classical
FT equation (\ref{TF}), as opposed to Eq. (\ref{density}) that relates the
electron density to the quantum mechanical Green's function.

The self-consistent solution in quantum transport or electronic structure
calculations is often found using a \textquotedblleft simple
mixing\textquotedblright\ method, when on each $m+1$ iteration step a small
part of a new potential is mixed with the old one (from the previous
iteration step), $V_{m+1}=(1-\epsilon )V_{m}+\epsilon \,V_{m+1}^{{}}$, $%
\epsilon $ being a small constant, $\sim 0.1-0.01$. It is typically needed $%
\sim 200-2000$ iteration steps to achieve our convergence criterium
\begin{equation}
\frac{|n^{m+1}-n^{m}|}{n^{m+1}+n^{m}}<10^{-5},  \label{convergence}
\end{equation}%
where $n^{m}=\int n^{m}(\mathbf{r})\,dy\,$ is the electron density in the
dot on $m$-th iteration step. In order to improve the convergence we employ
the modified Broyden's second method\cite{Singh} which allows us to reduce
drastically the number of iteration steps to $\sim 15-40.$ An input charge
density $n_{in}^{m+1}$ for $m+1$ iteration is constructed from the sets of
input and output densities ($n_{out}^{m},n_{in}^{m}$), from all $m$ previous
iterations
\begin{eqnarray}
n_{in}^{m+1} &=&n_{in}^{m}-B^{1}F^{m}-\sum_{j=2}^{m}U^{j}\left( V^{j}\right)
^{T}F^{m}  \label{Broyden} \\
F^{m} &=&n_{out}^{m}-n_{in}^{m},  \notag \\
U^{i} &=&-B^{1}\left( F^{i}-F^{i-1}\right) +n_{in}^{i}-n_{in}^{i-1}  \notag
\\
&&-\sum_{j=2}^{i-1}U^{j}\left( V^{j}\right) ^{T}\left( F^{i}-F^{i-1}\right) ,
\notag \\
\left( V^{i}\right) ^{T} &=&\frac{\left( F^{i}-F^{i-1}\right) ^{T}}{\left(
F^{i}-F^{i-1}\right) ^{T}\left( F^{i}-F^{i-1}\right) }.  \notag
\end{eqnarray}%
The initial guess $B^{1}$ is taken to be a small constant ($\sim 0.1-0.01$)
so that the input to the second iteration is effectively constructed using
the simple mixing. The scheme given by Eqs. (\ref{Broyden}) requires the
storage of relatively small number vectors which is thus more effective than
the original Broyden's second method.

\section{Results and discussion}

\subsection{A few electron open dot}

We calculate the magnetotransport of a split-gate open quantum with
following parameters representative for a typical experimental structure.
The 2DEG is buried at $b=60$ nm below the surface (the widths of the cap,
donor and spacer layers are 10 nm, 36 nm and 14 nm respectively), the donor
concentration is $0.6\cdot 10^{24}$ m$^{-3}$. The width of the
semi-infinitive leads is $w_{lead}=540$ nm, and the width of the
constrictions is $w_{qpc}=100$ nm (both the quantum point contacts are
identical), see Fig. \ref{f:structure} (b). The length of the quantum dot is
kept constant throughout the paper, $l_{dot}=160$ nm, while the width of the
dot is varied in the range $w_{dot}=170-440$ nm. The gate voltages applied
to the gates is $V_{lead}=-0.4$ V and $V_{qpc}=-0.44$ V. With these
parameters of the device there are 14 channels available for propagation in
the leads and the electron density in the center of the leads is $%
n_{lead}=1.6\cdot 10^{15}$ m$^{-2}$. The maximal electron density in the dot
(for $w_{dot}=440$ nm) is also $n_{dot}=1.6\cdot 10^{15}$ m$^{-2}.$ The
temperature is fixed at $T=0.2$ K for all results presented below.

In the following we discuss the open quantum dot with $N=1$ propagating
channel through both quantum point contacts. Note, that increasing the
number of propagating channels to $N=2,3$ does not qualitatively change the
results presented below. In order to set up the QPCs in the one-mode regime,
we grounded one of them and studied the conductance of the remaining QPC as
a function of the gate voltage $V_{qpc}$. The calculated conductance shows a
characteristic step-like dependence and we choose $V_{qpc}$ at the first
conductance plateau, namely, $V_{qpc}=-0.44$ V.

\begin{figure}[!tb]
\includegraphics[keepaspectratio,width=\columnwidth]{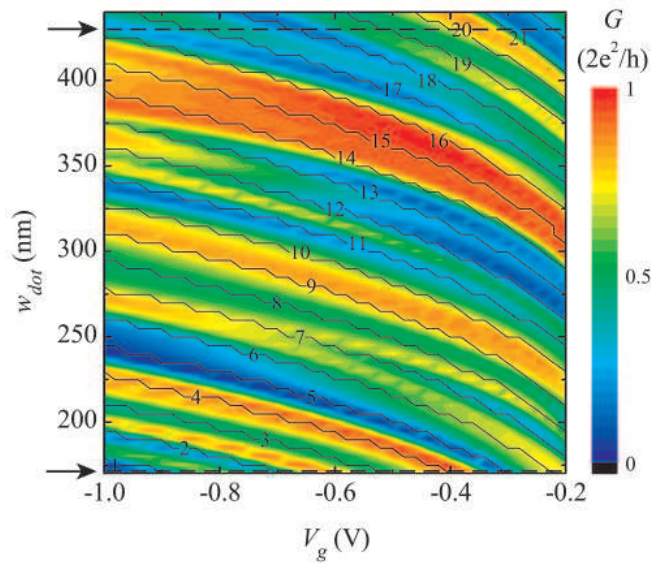}
\caption{(color online) The conductance of the open quantum dot as a
function of the width, $w_{dot}$, and the gate voltage $V_{g}$ calculated
within the Hartree approximation. Solid thin lines denote the number of
electrons. [Note that a zig-zag-type behavior of the electron number is an
artifact due to finite grid steps]. Arrows indicate $w_{dot}=170$ nm and $%
w_{dot}=430$ nm corresponding to two regimes, $\Delta \gg \Gamma$ and $%
\Delta \sim \Gamma$, discussed in Sec. IV.}
\label{f:2dplot}
\end{figure}

Figure \ref{f:2dplot} shows the color-scale plot of the conductance $G$ as a
function of the dot width $w_{dot}$ and the gate voltage $V_{g}$ (the
magnetic field is restricted to zero). The distinctive feature of the open
quantum dots is the oscillations of the conductance, in response to the
change of the geometrical size or the Fermi energy. (Various aspects of the
conductance oscillations in small and large dots have been the subject of
numerous experimental and theoretical works during the last decade\cite%
{Alhassid}). In the present study we concentrate on two dots with $%
w_{dot}=170$ nm and $w_{dot}=430$ nm (as indicated by arrows in Fig. \ref%
{f:2dplot}). As will be shown below, the first dot corresponds to the
transport regime when $\Delta \gg \Gamma ,$ whereas the second dot operates
in the regime $\Delta \sim \Gamma ,$ where $\Delta =\frac{2\pi \hbar ^{2}}{%
m^{\ast }S_{a}}$ is the mean level spacing separation in the dot with the
actual area $S_{a}$, and $\Gamma $ is the lead-induced broadening of the
resonant energy levels.

It is worth to note that a linear change of the gate voltage leads to a
nonlinear change of the effective dot size (If it were linear, the
conductance oscillations in Fig. \ref{f:2dplot} would exhibit a perfect
linear-stripe-type pattern). This is consistent with the experimental
results of Ref. \onlinecite{Andy_square} that show a deviation from the
linear dependence of the gate depletion distance as the gate voltage was
varied.

\subsection{Regime $\Delta \gg \Gamma $}

\begin{figure}[!tb]
\includegraphics[keepaspectratio,width=\columnwidth]{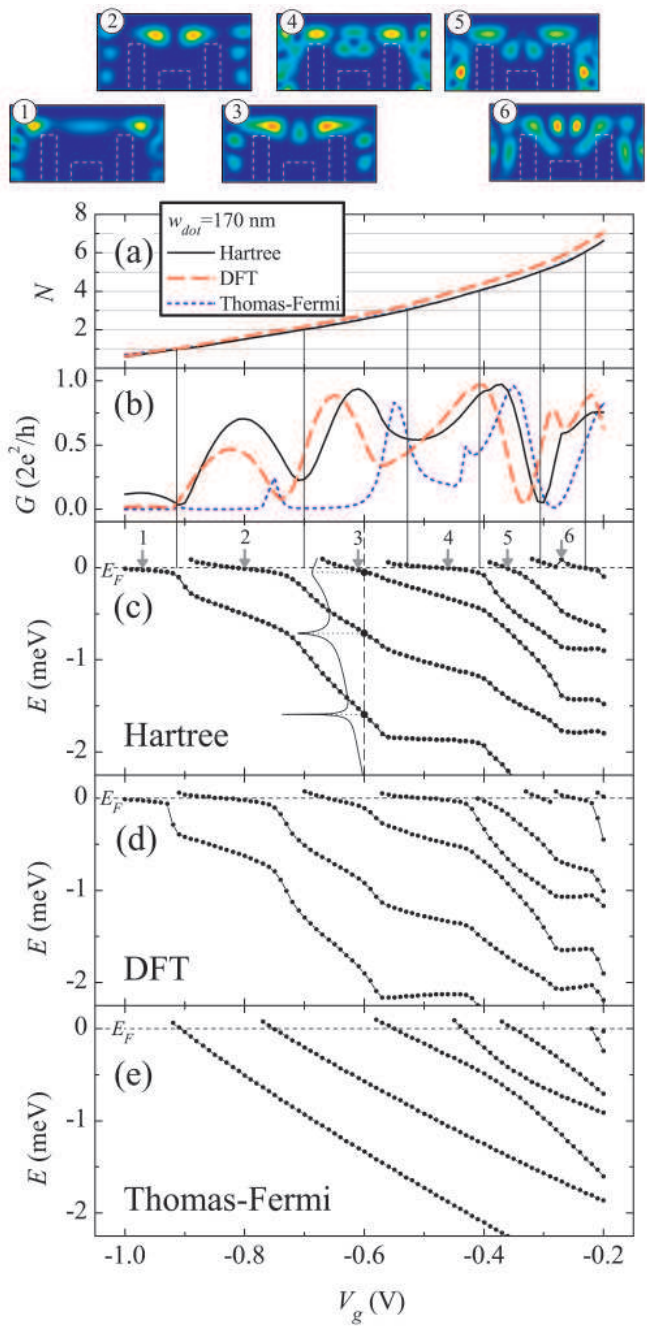}
\caption{(color online) (a) The number of electrons, (b) the conductance,
and (c)-(e) resonant energy structure in the few-electron open quantum dot
with $w_{dot}$=170 nm calculated within the Hartree, DFT and Thomas-Fermi
approximations. The top panel shows the electron probability amplitudes $%
\left|\Psi(x,y)\right|^{2}$ for the resonant energy levels marked by arrows
in (c). Inset in (c) shows the DOS for $V_{g}$=-0.6 V.}
\label{f:fewelectron}
\end{figure}

Figure \ref{f:fewelectron} shows the number of electrons in the few-electron
open quantum dot, the conductance and the peak-energy level position as a
function of the gate voltage $V_{g}$ for a quantum dot with $w_{dot}=170$ nm
calculated in the Hartree, DFT and TF approximations (The Hartree
approximation corresponds to disregarding of the exchange-correlation
potential in Eq. (\ref{V_tot}), $V_{xc}=0)$. The peak positions of the
resonant energy levels in the dot are extracted from the calculated DOS, as
illustrated in Fig. \ref{f:fewelectron} (c) for the case when the gate
voltage $V_{g}=-0.6$ V. The estimation of the mean level separation gives $%
\Delta \approx 0.6$ meV (for $V_{g}=-0.6$ V) which agrees quite well with
the actual level separation shown in Fig. \ref{f:fewelectron} (c) - (e).
[Note that within the given interval of variation of $V_{g}$, the actual dot
area changes and hence $\Delta $ varies as well]. An inspection of the DOS
shows that the the separation between the resonant levels are much larger
than the level broadening, $\Delta \gg \Gamma .$

All three approximations give very similar electron number $N$ in the dot as
a function of the gate voltage $V_{g}$. $N$ monotonically increases with
increase of $V_{g},$ which reflects the fact that electrons freely enter and
leave the dot in the open regime. However, the conductance calculated in the
Hartree and the DFT approximations exhibit rather similar behavior, whereas
the Thomas-Fermi conductance shows a very different gate voltage dependence.
The origin of this difference can be understood from the analysis of the
resonant level structure. \textit{The resonant energy levels calculated
within the Hartree and DFT approximations get pinned to the Fermi energy},
see Figs. \ref{f:fewelectron}(c) and (d) respectively. In stark contrast,
the energy level positions calculated in the Thomas-Fermi approximation
sweep past the Fermi level in a linear fashion when the applied voltage is
varied, Fig. \ref{f:fewelectron} (e).

The effect of level pinning is related to the screening properties of the
open quantum dot and the presence of the resonant structure in its DOS.
Indeed, in the vicinity of the resonances the DOS of the dot is enhanced
such that electrons with the energies close to $E_{F}$ (when $f_{FD}<1)$ can
easily screen the external potential. This leads to the metallic behavior of
the system when the electron density in the dot can be easily redistributed
to keep the potential constant. As a result, in the vicinity of a resonance
the system only weakly responds to the external perturbation (change of a
gate voltage, magnetic field, etc.), i.e. the resonant levels becomes pinned
to the Fermy energy. A comparison between the Hartree and the DFT
approximations indicate that the exchange-correlation interaction seems to
enhance the pinning, but an overall change is small (c.f. Figs. \ref%
{f:fewelectron} (c) and (d)). Thus, in the following we concentrate on the
Hartree approximation only. In contrast to the Hartree and the DFT
approaches, the effect of pinning is absent in the TF approximation. This is
because the total confining potential calculated within the TF approximation
does not capture the resonant structure of the DOS. Note that a modelling of
the magnetotransport in a quantum dot conceptually similar to our TF
approach was performed by Bird \textit{et al.}\cite{Bird}\textit{, }where
the confining potential for every given gate voltage was obtained as a
self-consistent solution of the Poisson equation.

Conductance oscillations in the open quantum dots can be related to
presence/absence of the resonant energy levels at $E_{F}$. Indeed, the first
three peaks in the Hartree conductance are attributed to the presence of
corresponding resonant levels, which is confirmed by the inspection of the
wave function $\left\vert \Psi (x,y)\right\vert ^{2}$ (by counting the
number of nodes of $\left\vert \Psi (x,y)\right\vert ^{2}$), Fig. \ref%
{f:fewelectron}. In turn, the dips indicate the absence of levels at $E_{F}$
and agree precisely with integer $N$ reflecting the fact that all the
available levels are below $E_{F}$ and thus are fully filled. Note that for
less negative gate voltages ($|V|\lesssim 0.3$ V) the separation between the
levels $\Delta $ becomes comparable to the broadening $\Gamma ,$ such that
the dips in the conductance are no longer correspond to the integer electron
number $N$ (see next section for details). It is also interesting to note
that the resonant state corresponding to the 5th eigenstate (5 nodes of the $%
\left\vert \Psi (x,y)\right\vert ^{2}$) is situated lower in energy than the
corresponding resonant state related to the 4th eigenstate. This is an
indication that 5th state couples with the leads more strongly that 4th state%
\cite{Zozoulenko_1997}.

The pinning of the resonant energy levels has an important effect on
transport in open quantum dots. In the considered transport regime, $\Delta
\gg \Gamma ,$ the conductance calculated within the one-electron (TF)
approximation exhibits distinct peaks separated by broad valleys of
essentially zero conductance. This reflects the structure of the
one-electron DOS where the resonant levels sweep past the Fermi level in a
linear fashion. In contrast, as a result of pinning, the DFT and Hartree
conductance shows much broader oscillations in comparison the one-electron
approach (Fig. \ref{f:fewelectron}(b)).



\subsection{Regime $\Delta \sim \Gamma $}

\begin{figure}[!tb]
\includegraphics[keepaspectratio,width=\columnwidth]{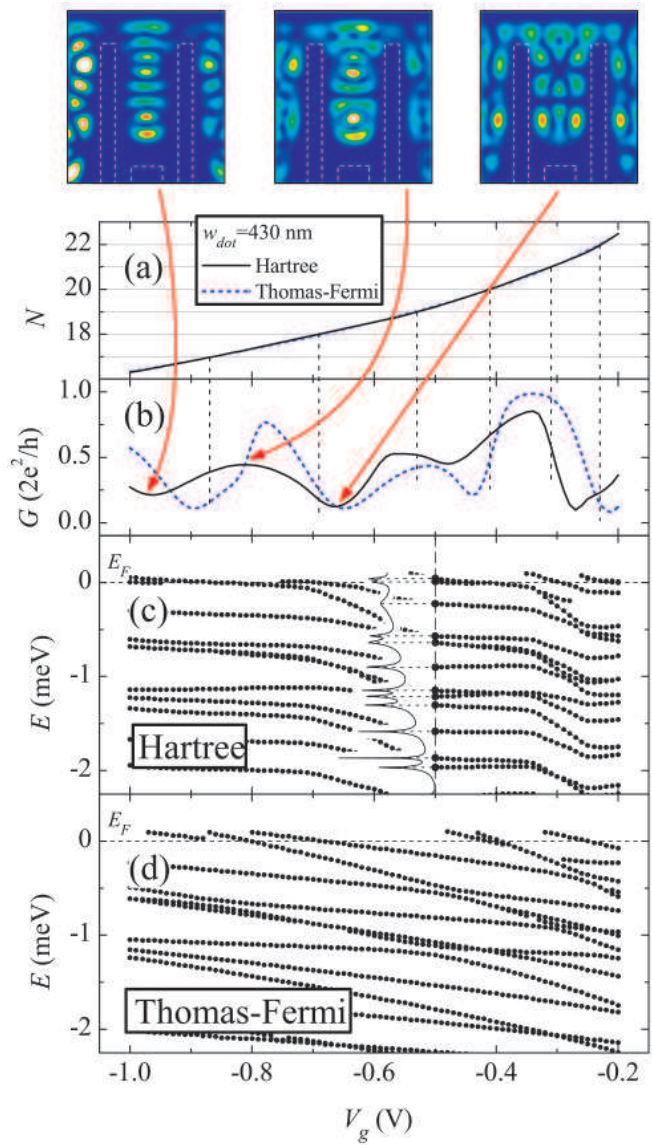}
\caption{(color online) (a) The number of electrons, (b) the conductance,
and (c)-(d) resonant energy structure in the few-electron open quantum dot
with $w_{dot}$=430 nm calculated within the Hartree and Thomas-Fermi
approximations. The electron probability amplitudes $\left|\Psi(x,y)%
\right|^{2}$ are shown for some representative $V_{g}$ (top panel). Inset in
(c) shows the DOS for $V_{g}$=-0.5 V.}
\label{f:sevelectron}
\end{figure}

When an effective size of a quantum dot increases, the mean level spacing
separation $\Delta $ decreases. For the quantum dot of the size $w_{dot}=430$
the estimation of the mean level separation gives $\Delta \approx 0.15$ meV
(for $V_{g}=-0.6$ V) which agrees quite well with the actual level
separation shown in Fig. \ref{f:sevelectron} (c) - (d). An inspection of the
DOS shows that for this dot the spacing between neighboring levels is
comparable with the level broadening, $\Delta \sim \Gamma ,$ see Fig. \ref%
{f:sevelectron}(c). [Note, that the level broadening is controlled by
coupling to the leads and does not depend on the dot size.] Because
neighboring levels start to overlap, there is always one or several energy
states at $E_{F}$ mediating transport through the dot. The effect of the
pinning of the energy levels in the Hartree calculations (as well as its
absence in the TF calculations) is also clearly seen in this regime as well.
However, in contrast to the regime $\Delta \gg \Gamma ,$ the dips in the
conductance can no longer be related to the integer number of electrons in
the dot. Instead, it is the interference between states at the entrance and
exit of the open quantum that determines the dependence $G=G(V_{g})$\cite%
{Zozoulenko_1997}. Our preliminary results for the conductance of even
larger dots (containing hundreds of electrons) outline different features of
the TF and Hartree conductances that are manifest themselves in the
amplitude and the broadening of the conductance peaks. A detailed analysis
of the statistics of the conductance oscillations is however outside the
scope of the present work and will be deferred to future publications.

\subsection{Effect of magnetic field}

\begin{figure}[!tb]
\includegraphics[keepaspectratio,width=\columnwidth]{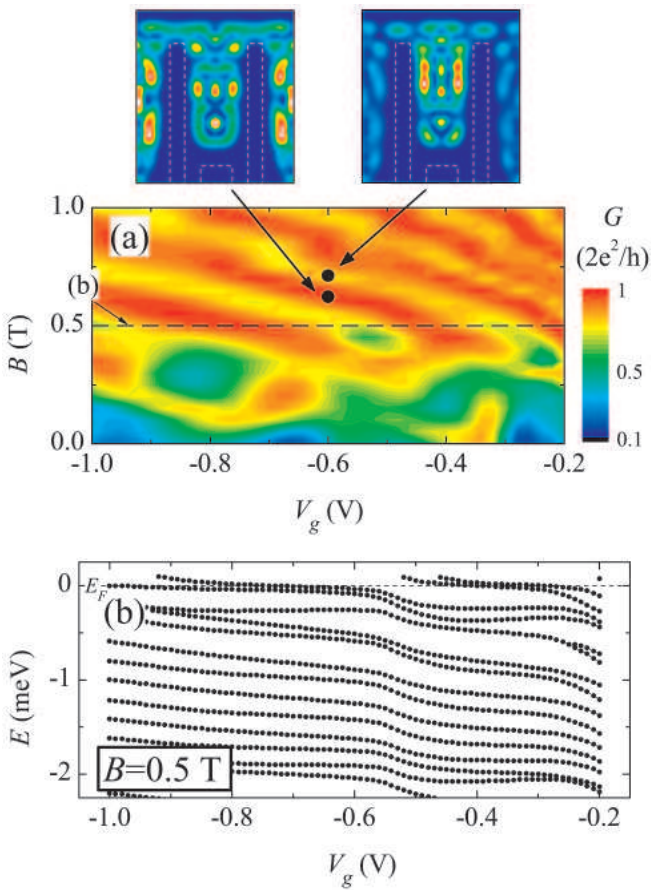}
\caption{(color online) (a) The conductance of the open quantum dot as a
function of the magnetic field $B$ and gate voltage $V_{g}$ for the open
quantum dot of the width $w_{dot}$=430 nm calculated within the Hartree
approximation. (b) The energy structure for $B=0.5$ T calculated within the
Hartree approximation. The top panel shows the electron probability
amplitudes $\left|\Psi(x,y)\right|^{2}$ (top) for two representative
magnetic fields.}
\label{f:B}
\end{figure}

In a sufficiently high magnetic field the electron transport takes place by
the edge states with a characteristic dimension of the order of the magnetic
length $l_{B}=\sqrt{\hbar /eB}$. In the edge state transport regime
backscattering on the potential defining the quantum dot decreases and, for
a large enough $B$, electrons pass through the device with the transmission
close to unity. Transport in such a regime is referred to as adiabatic. For
the open quantum dot of $w_{dot}=430$ nm, transition to adiabatic
propagation takes place at about $B\approx 0.5$ T, see Fig. \ref{f:B}(a).
The conductance for $B\gtrsim 0.5$ T shows pronounced oscillations due to
the Aharonov-Bohm interference. When the magnetic field changes such that
the total magnetic flux $\Phi =BS$ through the dot modifies by the one flux
quantum $\phi _{0}=h/e$, the conductance demonstrates periodic oscillations
with the period $\Delta B=\phi _{0}/S$ ($S$ is the characteristic area of
the dot). Using the actual dot area $S_{a}$ we get $\Delta B=0.11$ T, which
is nearly twice less than extracted from Fig. \ref{f:B}(a), where $\Delta
B=0.25$ T. The discrepancy can be related to a finite extent of the edge
state circulating inside the dot ($l_{B}\approx 35$ nm for $B=0.5$). As a
result, the area enclosed by the edge state is much smaller than the
geometrical area of the dot.

The resonant energy structure is modified substantially when a magnetic
field is applied, c.f. Fig. \ref{f:B}(b) for $B=0.5$ T and Fig. \ref%
{f:sevelectron} (c) for $B=0$ T. The resonant levels exhibit almost equal
separation which can be related to the well-known Darvin-Fock-type energy
spectrum formation for the corresponding closed dot \cite{Davies_book}. The
distinguished feature of the energy level structure is much stronger pinning
of the resonant levels to $E_{F}$ that persists over larger intervals of $%
V_{g}$ in comparison to the $B=0$ case. Stronger pinning can be attributed
to the enhanced screening efficiency because of the increased localization
of the wave function for the case of nonzero magnetic field. As we mentioned
in the introduction the strong pinning of the resonant energy levels in the
presence of the magnetic field can have a profound effect on thransport
properties of various devices operating in the edge state transport regime
including the Mach-Zender\cite{Mach-Zender} and the Laughlin\cite{Goldman}
interferometers as well as antidot devices\cite{QuantComp,adot,Karakurt,APL}.

To conclude this section, it is worth mentioning that another manifestation
of the screening in the edge state regime is the well-known effect of
formation of the compressible and incompressible strips near the structure
boundary\cite{Chklovskii}.

\subsection{Open vs closed system}

In modelling quantum mechanical transport in quantum dots and related
systems one often uses an approximation where an inherently open system is
replaces by a corresponding large, but closed one, see e.g. \cite{Jovanovic}%
. In this section we critically examine such an approximation. In
particular, we address a question whether a conductance calculated in such a
way coincides with the conductance of a truly open system, and whether the
pinning of the resonant energy levels is survived or not.
\begin{figure}[tb]
\includegraphics[keepaspectratio,width=\columnwidth]{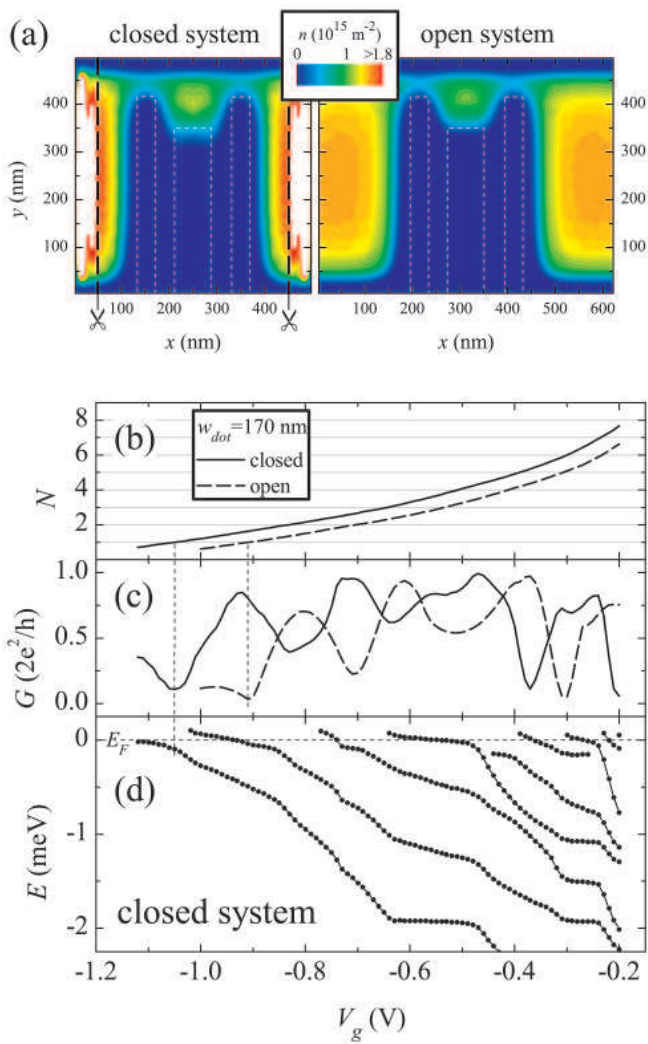}
\caption{(color online) (a) The representative self-consistent charge
densities for the closed and open systems. The thick dashed lines show the
cuts for the transport calculations. (b) The electron number, (c) the
conductance and (d) the peak energy-level position calculated within the
Hartree approximation in the closed-system approximation. The dashed lines
denote the result for the open system (the same as in Fig. \protect\ref%
{f:fewelectron}). The width of the quantum dot is $w_{dot}$=170 nm.}
\label{f:closed}
\end{figure}
For modelling of the closed system, we replace the semi-infinitive leads by
the potential walls of an infinitive height such that the solution of the
Schrodinger equation reduces to the eigenproblem
\begin{equation}
H\psi =E\psi ,  \label{eigenproblem}
\end{equation}%
where $E$ and $\psi $ are discrete sets of eigenvalues and eigenvectors, and
the Hamiltonian $H$ is given by Eq. (\ref{Hamiltonian}). We solve Eq. (\ref%
{eigenproblem}) self-consistently performing the fast Fourier transformation
from the coordinate into the energy space, which greatly reduces a
computational cost.

In our calculations we fix the Fermi energy such that the charge density in
the system is given as
\begin{equation}
n(\mathbf{r})=\sum_{i}\psi _{i}(\mathbf{r})f(E_{i}-E_{F}).  \label{density2}
\end{equation}%
This assumption leads to a noninteger electron number in a system, but we
apriory construct the closed system resembling the open one as much as
possible. Note also, that because the total number of electrons $N_{tot}\gg
1,$ the effect of the non-integer $N_{tot}$ on the total potential is
practically negligible.

In order to calculate the conductance of the system at hand, we cut off
slices in the vicinity of the boundaries as illustrated in Fig. \ref%
{f:closed}(a) and then add homogeneous semi-infinite leads with the
potential that matches the potential of the boundary slices. Finally, we
solve a scattering problem for this given potential using the recursive
Greens function technique\cite{Zozoulenko_1996}.

Figures \ref{f:closed}(b)-(d) show the electron number in the dot $N$, the
conductance $G$ and the resonant energy structure within the Hartree
approximation. The comparison to the corresponding results for the open
system shows that the closed-system approximation reproduces all the results
not only qualitatively but rather quantitatively, c.f. Fig. \ref%
{f:fewelectron}.The only difference is the shift along $V_{g}$-axis which is
simply related to the fact that the Hartree potential for the case of the
closed system, in contrast to the open one, does not include a contribution
from the semi-infinite leads. Figure \ref{f:closed}(d) reveals that the
pinning of resonant energy levels to the Fermi energy is present in the
closed system-approximation as well. We thus conclude that this
approximation might be used for the modelling of the transport properties
and the resonant energy level structure of the corresponding open system. We
however should note that with the present approximation we could not satisfy
the convergence criterium (\ref{convergence}) that we routinely use for the
calculation of the conductance in the open systems as described in Sec. III.

\section{Conclusion}

We have developed an approach for full quantum mechanical many-body
magnetotransport calculations in open systems that starts from the
lithographical layout of the device and does not include phenomenological
parameters like coupling strengths, charging constants etc. The whole
device, including semi-infinitive leads, is treated on the same footing
(i.e. the electron-electron interaction is accounted for in both the leads
as well as in the dot region). The many-body effects are included within the
DFT and the Hartree approximations.

The developed method was applied to calculate the conductance through an
open quantum dot. The main finding of the present paper is the effect of
pinning of the resonant levels to the Fermi energy due to the enhanced
screening. Our results represent a significant departure from a conventional
picture where a variation of external parameters (such as a gate voltage,
magnetic field, etc.) causes the successive dot states to sweep past the
Fermi level in a linear fashion. We instead demonstrate highly nonlinear
behavior of the resonant levels in the vicinity of the Fermi energy. We show
that the pinning effect is absent in a one-electron (Thomas-Fermy)
approximation because in this case the self-consistent potential does not
account for the resonant structure of the DOS in the dot. The pinning of the
resonant levels in open quantum dots leads to the broadening of the
conduction oscillations in comparison to the one electron picture. It
remains to be seen whether accounting for this effect might shed new light
on the interpretation of the conductance oscillation statistics in open
quantum dots.

The pinning of the resonant levels becomes much more pronounced in the
presence of the perpendicular magnetic field. This can be attributed to the
enhanced screening efficiency because of the increased localization of the
wave function. The strong pinning of the resonant energy levels in the
presence of magnetic field can have a profound effect on transport
properties of various devices operating in the edge state transport regime
including Mach-Zender\cite{Mach-Zender} and Laughlin\cite{Goldman}
interferometers as well as antidot devices\cite{QuantComp,adot,Karakurt,APL}.

We should stress that the pinning effect predicted in this paper is not
specific to the considered material system (GaAs/AlGaAs heterostructure) and
is expected to hold in any two-dimensional system in open transport regime
(e.g. Si inversion layer structures, etc.).

Finally, in the present paper we critically examined an approximation used
in modelling of quantum mechanical transport in quantum dots and related
systems when an inherently open system is replaces by a corresponding large,
but closed one.

In the present study we have limited ourselves to the case of spinless
electrons. Work is in progress to include the effect of the spin in order to
revisit the effect of spin splitting recently observed in open quantum dots%
\cite{EurophLett,Evaldsson}.

\bigskip


\begin{acknowledgments}
S. I. acknowledges financial support from the Swedish Institute and the EU
network SINANO. Numerical calculations were performed in part using the
facilities of the National Supercomputer Center, Link\"{o}ping, Sweden.
\end{acknowledgments}

\end{document}